\documentclass[aps,prc,showpacs,amssymb,amsmath,amsfonts,nofootinbib,floatfix]{revtex4-1}
\usepackage{graphicx}

\makeatletter
\def\p@subsection{}
\def\p@subsubsection{}
\makeatother
\usepackage{natbib}
\usepackage[usenames]{color}
\usepackage{epsfig}
\usepackage{epstopdf}
\usepackage{amssymb,amsmath}
\usepackage{amsmath}
\usepackage{subfig}
\usepackage{epstopdf}
\usepackage{enumitem}

\definecolor{grey}{rgb}{0.9,0.9,0.9}
\definecolor{black}{rgb}{0,0,0}

\newcommand{\di}{i}

\newcommand{\be}{\begin{eqnarray}}
\newcommand{\ee}{\end{eqnarray}}
\newcommand{\bc}{\begin{center}}
\newcommand{\ec}{\end{center}}
\newcommand{\beq}{\begin{eqnarray}}
\newcommand{\eea}{\end{eqnarray}}

\begin{document}

\title{From Experimental Data to Pole Parameters in a Direct Way \\
{\normalsize (Angle Dependent Continuum Ambiguity and  Laurent + Pietarinen Expansion)}
}

\author{ A. \v{S}varc\,$^{1}$ \\
 Y. Wunderlich\,$^2$, H. Osmanovi\'{c}\,$^3$ \\
  M. Had\v{z}imehmedovi\'{c}$\,^3$, R. Omerovi\'{c}$\,^3$, J. Stahov$\,^3$, \\  V. Kashevarov$\,^4$, K. Nikonov$\,^4$, M. Ostrick$\,^4$, L. Tiator$\,^4$  \\
R. Workman$^5$   }
\email{Corresponding author: alfred.svarc@irb.hr}
\affiliation{$\,^1$ Rudjer Bo\v{s}kovi\'{c} Institute, Bijeni\v{c}ka cesta 54,
                 P.O. Box 180, 10002 Zagreb, Croatia}
\affiliation{$\,^2$ Helmholtz-Institut f\"{u}r Strahlen- und Kernphysik der Universit\"{o}t Bonn, Nussallee 14-16, 53115 Bonn, Germany }
\affiliation{$\,^3$ University of Tuzla, Faculty of Natural Sciences and Mathematics, Univerzitetska 4, 75000 Tuzla, Bosnia and Herzegovina}
\affiliation{$\,^4$ Institut f\"{u}r Kernphysik, Universit\"{a}t Mainz, D-55099 Mainz, Germany}
\affiliation{$\,^5$ Data Analysis Center at the Institute for Nuclear Studies, Department of Physics, The George Washington University, Washington, D.C. 20052, USA}

\vspace{5cm}
\date{\today}

\begin{abstract}

Unconstrained partial-wave amplitudes obtained at discrete energies from fits to complete sets of eight independent observables which are required to uniquely reconstruct reaction amplitudes
do not vary smoothly with energy, and are in principle non-unique. We demonstrate how this behavior can be ascribed to the continuum
ambiguity. Starting from the spinless scattering case, we demonstrate how an unknown
overall phase depending on energy and angle mixes the structures seen in the associated partial-wave amplitudes making the partial wave decomposition non-unique, and illustrate it on a simple toy model.
We then apply these principles to pseudo-scalar meson photoproduction and show that the non-uniqueness effect can be
removed through a phase rotation generating \mbox{"up-to-a-phase"} unique set of SE partial wave amplitudes. Extracting pole positions from partial wave amplitudes is the next step.
Up to now, there was no reliable way to extract pole parameters from SE partial waves, but a new and simple single-channel method (Laurent + Pietarinen expansion) applicable for continuous and discrete data has been recently developed. It is based on applying the Laurent decomposition of partial wave amplitude, and expanding the non-resonant background into a power series of a conformal-mapping, quickly converging power series obtaining the simplest analytic function with well-defined partial wave analytic properties which fits the input.  The generalization of this method to multi- channel case is also developed and presented.
Unifying both methods in succession, one constructs a model independent procedure to extract pole parameters directly from experimental data without referring to any theoretical model.

\end{abstract}

\maketitle
\section{Introduction}

Partial-wave analysis, a textbook method to identify resonances and attribute them quantum numbers, is a standard procedure used to analyze a wide class of experimental data
(see for instance ref. \cite{MartinSpearman}).
This paper deals with a very general interplay of partial wave decomposition  and invariance with respect to a general symmetry of energy and angle dependent phase rotation, and  is therefore of wide interest for all scientists which apply the method of partial wave decomposition for their data analysis. At the same time this paper points out, but also solves a critical, but up to now unknown problem of uniqueness of partial wave decomposition and has a profound impact on research in the field of nucleon resonances.

In the quest for a unique set of amplitudes,
experimental programs have attempted to measure complete sets of observables needed to perform the unambiguous  reaction-amplitude reconstruction.  On the other hand, it has also been well known that single-channel physical observables remain invariant with respect to a general energy- and angle-dependent phase rotation which turns out to introduce the non-uniqueness into the amplitude reconstruction.
This invariance is the so-called continuum ambiguity, and has been extensively analyzed in mid-70s through
the mid-80s \cite{Atk73,Bow75,Atk85}.  The interconnection of the two has never been discussed as  most studies were made in the context of $\pi N$ elastic scattering where the
optical theorem and an application of elastic unitarity practically eliminate the continuum ambiguity as a source of non-uniqueness.  In the inelastic domain, up to now, the main attention has been paid only to handling angle independent phase rotations on the level of partial waves where the analytic structure  remains untouched \cite{bnga,Tiator:2011pw,Sandorfi2011}.  We show that, contrary to well-behaved energy dependent invariance, angle dependent phase rotation mixes partial waves, and changes their analytic structure causing partial wave decomposition to be non-unique. This mixing directly influences the present identification of resonance quantum numbers. The discussion of angle dependent phase rotation was always scarce, and to our knowledge, mentioned only once, and this only "in passing" in old ref.~\cite{Oma81}.
Some other related studies on phase ambiguities were performed \cite{Dean,Keaton}, but concentrated on cases without a complete set
of experimental data. Programmatic studies of photoproduction experiments at Jefferson Lab, Mainz and Bonn are now producing the data required to do complete experiments, in terms of either helicity amplitudes or multipoles, motivating a reexamination of
the ambiguities associated with multipole analyses.

In this paper we show that, contrary to well-behaved energy-dependent phase rotations, angle-dependent phase rotations
mix partial waves, changing their analytic structure and producing partial wave decompositions which are
non-unique. This directly influences the present identification of resonance quantum numbers.   We then explore the effect of angle-dependent rotations in a
case of practical interest, the single-energy analysis of a complete set of eta photoproduction pseudo-data generated from a known model by comparing produced multipoles with the known model-starting values. We confirm that the obtained result is non-unique.  As a solution of this problem we offer a method how a partial wave decomposition can be made up-to-a-phase unique by replacing the undetermined and discontinuous energy and angle dependent phase with an arbitrary but continuous chosen value. This is by no means a restriction of generality as all other solutions with different amplitude phases can be obtained directly from the obtained solution by a phase rotation on the level of reaction amplitudes, changing the arbitrary to any particularly
 given phase. All solutions remain equivalent as far as fitting the data is concerned. For the true quantum number identification, however, one still has to identify the unknown continuum ambiguity phase because each phase gives its own set of quantum numbers. In fully unconstrained single-channel analyses, quantum numbers remain undetermined.

 Up to now, there was no reliable way to extract pole parameters from so obtained SE partial waves, but a new and simple single-channel method (Laurent + Pietarinen expansion) applicable for continuous and discrete data has been recently developed \cite{Svarc2013,Svarc2014,Svarc2014a}. It is based on applying the Laurent decomposition of partial wave amplitude, and expanding the non-resonant background into a power series of a conformal-mapping, quickly converging power series obtaining the simplest analytic function with well-defined partial wave analytic properties which fits the input. The method is particularly useful to analyse partial wave data obtained directly from experiment because it works with minimal theoretical bias since it avoids constructing and solving elaborate theoretical models, and fitting the final parameters to the input, what is the standard procedure now.  The generalization of this method to multi- channel case is also developed and presented.

 Unifying both methods in succession, one constructs a model independent procedure to extract pole parameters directly from experimental data without referring to any theoretical model.

\section{Formalism}
Let us recall that observables in single-channel reactions are given as a sum of products involving one
(helicity or transversity) amplitude with the complex conjugate of another, so that the general form of
any observable is ${\cal O}=f˙(H_k \cdot H_l^*)$, where $f$ is a known, well-defined real function.
The direct consequence is that any observable is invariant with respect to the following simultaneous
phase transformation of all amplitudes:
\be
H_k (W,\theta) \rightarrow  \tilde{H_k} (W,\theta)& = & e^{\, i \, \,  \phi(W,\theta) } \cdot H_k (W,\theta) \nonumber \\
 {\rm for \, \, all} \, \, k & = & 1,\cdots,n
\ee
\noindent
where n is the number of spin degrees of freedom (n=1 for the 1-dim toy model, n=2 for pi-N scattering
and n=4 for pseudoscalar meson photoproduction), and $\phi(W,\theta)$ is an arbitrary, real function which is the same for all contributing amplitudes.

As resonance properties are usually the goal of such studies, and these are identified with poles of the
partial-wave (or multipole) amplitudes, we must analyze the influence of the continuum ambiguity not
upon helicity or transversity amplitudes, but upon their partial wave decompositions.
To simplify the study we introduce partial waves in a simplified version than those found in
Ref. \cite{Tiator:2011pw}:
\be \label{Eq:PW}
A (W,\theta) & = & \sum^{\infty}_{\ell=0} (2 \ell + 1) A_\ell(W) P_{\ell}(\cos\theta)
\ee
\noindent
where $A(W,\theta)$ is a generic notation for any amplitude $H_k(W,\theta)$, $k=1, \cdots n$.
The complete set of observables remains unchanged when we make the following transformation:
\be \label{Eq:PWrot}
A (W,\theta) \rightarrow \tilde{A} (W,\theta)& = & e^{\, \di \, \,  \phi(W,\theta) } \nonumber \\
& \hspace*{10pt} \times &  \sum^{\infty}_{\ell=0} (2 \ell + 1) A_\ell(W) P_{\ell}(\cos\theta) \nonumber \\
\tilde{A} (W,\theta) & = &  \sum^{\infty}_{\ell=0} (2 \ell + 1) \tilde{A}_\ell(W)P_{\ell}(\cos\theta)
\ee
\noindent
We are interested in rotated partial wave amplitudes $\tilde{A}_\ell(W)$, defined by Eq.(\ref{Eq:PWrot}),
and are free to introduce the Legendre decomposition of an exponential function as:
\be \label{Eq:Phaseexpansion}
e^{\, \di \, \,  \phi(W,\theta) } &=& \sum^{\infty}_{\ell=0} L_\ell(W)  P_{\ell}(\cos\theta).
\ee
After some manipulation of the product
$P_\ell(x) P_k(x)$
(see refs.~\cite{Dougall1952,Wunderlich2017}~for details of the summation rearrangement) we obtain:
\be \label{Eq:mixing}
\tilde{A}_\ell(W) &=& \sum_{\ell'=0}^{\infty} L_{\ell'}(W) \, \, \, \cdot \sum_{m=|\ell'-\ell|}^{\ell'+\ell}\langle \ell',0;\ell,0|m,0 \rangle ^2\, \, A_{m}(W) \nonumber \\
\ee
where $\langle \ell',0;\ell,0|m,0 \rangle $ is a standard Clebsch-Gordan coefficient.
\\ \indent
To get a better insight into the mechanism of multipole mixing, let us expand Eq.~(\ref{Eq:mixing}) in terms of phase-rotation Legendre coefficients $L_{\ell'}(W)$: \\
\begin{align} \label{Eq:mixing-expanded}
 \tilde{A}_{0} (W) &= {\bf L_{0} (W) A_{0} (W) } + L_{1} (W) A_{1} (W) + L_{2} (W) A_{2} (W) + \ldots  \mathrm{,} \\
 \tilde{A}_{1} (W) &= {\bf L_{0} (W) A_{1} (W) } + L_{1} (W) \left[\frac{1}{3}A_{0} (W) + \frac{2}{3} A_{2} (W) \right] + L_{2} (W) \left[\frac{2}{5} A_{1} (W) + \frac{3}{5} A_{3} (W) \right] + \ldots \mathrm{,} \nonumber \\
 \tilde{A}_{2} (W) &= {\bf L_{0} (W) A_{2} (W) } + L_{1} (W) \left[\frac{2}{5} A_{1} (W) +\frac{3}{5} A_{3} (W) \right] + L_{2} (W) \left[\frac{1}{5} A_{0} (W) + \frac{2}{7} A_{2} (W) + \frac{18}{35} A_{4} (W) \right] + \ldots \mathrm{.}  \nonumber \\
 \vdots \nonumber
\end{align}
The consequence of Eqs.~(\ref{Eq:mixing})~and~(\ref{Eq:mixing-expanded}) is that angular-dependent phase rotations
mix multipoles.
\\ \\ \noindent
\underline{\emph{Conclusion:}}
\\ \\ \indent
Without fixing the free continuum ambiguity phase $\phi(W,\theta)$,
the partial wave decomposition $A_\ell (W)$ defined in Eq.~(\ref{Eq:PW}) is non-unique. Partial waves get mixed, and  identification of resonance quantum numbers might be changed.
To compare different partial-wave analyses, it is  essential to match the continuum ambiguity phase;
otherwise the mixing of multipoles is yet another, uncontrolled, source of systematic errors. Observe that this phase rotation does not create new pole positions, but just reshuffles the existing ones among several partial waves.

\section{Toy model:}

\begin{figure*}[h]
\begin{center}
\includegraphics[width=0.8\textwidth]{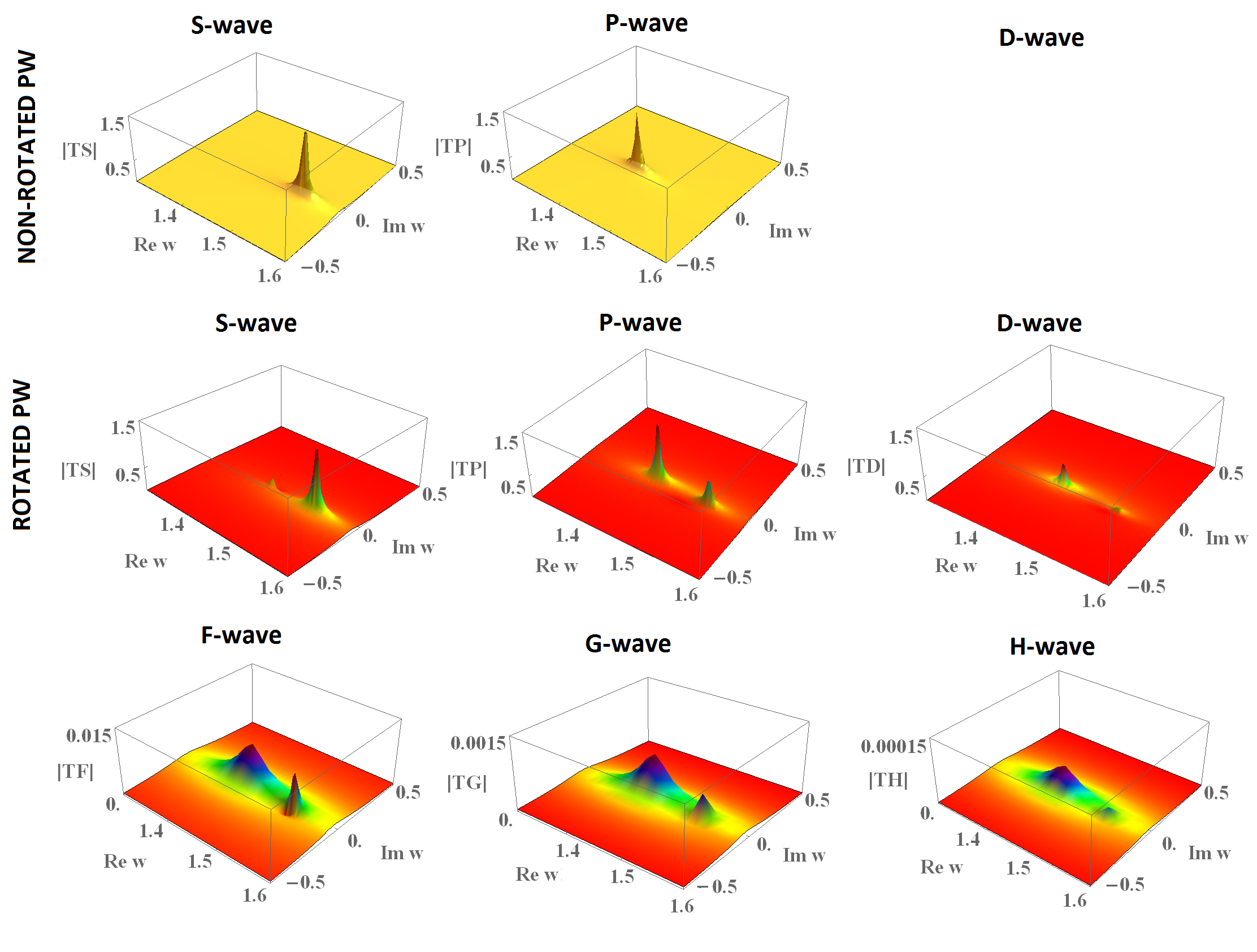}\vspace{-3mm}
\caption{\label{Toy model}(Color online) Toy model poles for linear rotation. }
\end{center}
\end{figure*}

To better illustrate the effect of mixing partial waves in the photo-production channel, we construct a simple toy model consisting of two partial waves,
with one resonance per partial wave.
\be
A(W,\theta) &=& T_S(W) + x \, \, T_P (W) \nonumber \\ \nonumber \\
T_{S,P}(W) &=&  \frac{a_{S,P}}{M_{S,P} - \di \, \Gamma_{S,P}/2-W} \\
     x & = & \cos \theta
\ee
where
\begin{center}
\begin{tabular}{ccc}
$a_S = (0.005 + \di \, 0.004$) GeV ; \hspace*{0.7cm}    & $M_S = 1.535 $ GeV; \hspace*{0.7cm} &  $\Gamma_S = 0.15$ GeV  \\
$a_P = (0.004 + \di \, 0.003$)  GeV ; \hspace*{0.7cm}   & $M_P = 1.440 $ GeV; \hspace*{0.7cm} &  $\Gamma_P = 0.10$  GeV.\\ \\
\end{tabular}
\end{center}
We take a very simple linear rotation acting upon the full amplitude $A(W,\theta)$:
\be \label{Toymodelrotation}
\mathcal{R}(x) &=& e^{ \di \, (2. + 0.5 \, x)},
\ee
and compare the partial-wave decompositions of the non-rotated and rotated amplitudes.
In Fig.~\ref{Toy model} we show the result.

\newpage
On the basis of Eqs.~(\ref{Eq:mixing})~and~(\ref{Eq:mixing-expanded}) applied to the toy model,
we observe several important features of the angle-dependent phase rotations:
\begin{enumerate}
\item The Legendre expansion of the angle-dependent phase rotation, given by
Eq.~(\ref{Eq:Phaseexpansion}), has, in principle, an infinite number of terms.
However, in cases of simpler rotations, higher order terms rapidly become very small. In this case the partial wave mixing is limited only to the neighbouring
partial waves (see Eq.~(\ref{Eq:mixing-expanded})).\footnote{For more complicated rotations, the ones where associated Legendre coefficients
extend to infinity, the situation gets more unstable as the resonance with the very high angular momentum can be
intermixed even into usually dominating S-wave. So, association of quantum numbers becomes even more questionable.}
\item The lowest rotation coefficient $L_0(W)$ transforms  $A_\ell(W)$ to  $\tilde{A}_\ell(W)$.
        Hence, for angle-independent rotations only the 0-th order Legendre coefficient survives, and
such a rotation does not mix multipoles.
\item Even the simplest quickly converging rotations, with only a few Legendre coefficients
significantly greater than zero, will mix multipoles. The degree of mixing rises with the complexity of
the angular rotation.
\item As explicitly seen in Eq.~(\ref{Eq:mixing-expanded}) under the influence of any phase rotation  only existing poles gets redistributed. So, through the influence of the angle-dependent phase rotation, no new pole positions are created,
but resonances with the same pole positions appear in different partial waves making the quantum number identification difficult. The only way out is to identify the genuine
continuum ambiguity phase where the partial-wave matrix gets diagonal, and only one resonance is associated with one partial wave.\footnote{We can understand all partial waves as being a matrix elements of the partial-wave-operator defined on the space of Legendre polynomials which form an orthonormal basis. Without knowing the genuine continuum ambiguity phase that operator is non-diagonal, and same resonances appear in different matrix-element positions hence having different quantum numbers. However, we can always diagonalize the partial-wave-operator (find such a basis of Legendre polynomials where this operator is diagonal), and in this base only one partial wave is associated with one Legendre polynomial defining the true resonance quantum numbers. Let us add that this new base where the partial wave operator is diagonal corresponds to the true continuum ambiguity phase.  }
\end{enumerate}

Another interesting effect can be observed. The toy-model $A(W,\theta)$ truncated at $L=1$
leads to 'data' for the differential cross section whose Legendre-expansion
\begin{equation}
\sigma_{0} (W,\theta) = \left|A(W,\theta)\right|^{2} = \sum_{n=0}^{2} a_{n} (W) P_n (\cos \theta) \mathrm{,} \label{eq:DCSLegendreExpansion}
\end{equation}
would be precisely truncated at $2L=2\ast1=2$, i.e. all Legendre coefficients up
to and including $a_{2}$ would be non-vanishing, with all higher coefficients,
i.e. $a_{3}$ and above, being  precisely zero.
If one then defined a new amplitude, with the angle-dependent phase (\ref{Toymodelrotation})
multiplied into the original
truncated amplitude, we would end up with an, in principle, infinite partial-wave expansion
for the rotated amplitude (which, however, converges quite well
for the higher $\ell$).
In fact, the rotation (\ref{Toymodelrotation}) itself turns out to
be truncated at $L^{\prime}=5$ to a very good approximation, since all of its
Legendre coefficients $L^{\prime}{\geq6}$ have moduli within the range of $10^{-7}$ or less.

Now, one might naively expect the Legendre-expansion of the data
(\ref{eq:DCSLegendreExpansion}) to show non-vanishing Legendre-coefficients up to and including
$2(L + L^{\prime}) = 2*(1 + 5) = 12$, once the rotated amplitude has been inserted.
This, however, is is not what happens. Since the phase (\ref{Toymodelrotation}) has
modulus $1$ for all $x$, it will leave the cross section invariant. The data simply do not change.
Since the original data are truncated at $2\ast1 = 2$, by construction, the rotated data will be as well.
However, the rotated amplitude itself
is manifestly not truncated. Furthermore, higher Legendre-coefficients $a_{\, \ell \geq 3}$
are bilinear hermitean forms, depending on the rotated partial waves.
All this implies that a cancellation-effect has to set all 'rotated' Legendre coefficients
$a_{3}$,...,$a_{12}$ to zero. The rotation (\ref{Toymodelrotation}) has generated higher partial waves in such a
'finely tuned' way that this is indeed possible.
\clearpage
\section{Pseudo-data}
Having explored a simple toy model, we next consider a more realistic analysis of precise pseudo-data generated
from an existing model for this chosen reaction. Just for the convenience of the reader let us summarize the essence of pseudo-data method.

Using pseudo-data for testing new procedures is a textbook method used a lot in our field (see for instance ref~\cite{Pseudo-data}).  In general we use it to test whether a certain procedure is correct by performing it on a set of data artificially created from a known source, and for which we know the exact answer in advance.  In this paper  we use it to test whether phase rotation can be used to achieve uniqueness of unconstrained SE PWA. We construct a complete set of observables out of well know model-partial-waves, so we know what should we get if the recommended method is correct. Then we perform the procedure, and if we reproduce the input-partial-waves uniquely, we know that our method is valid. Observe that it is completely unimportant what did we take as a generating model as long as our procedure is consistently implemented,  so we shall not discuss the features of the model used.

\begin{figure}[h!]
\begin{center}
\includegraphics[width=0.45\textwidth]{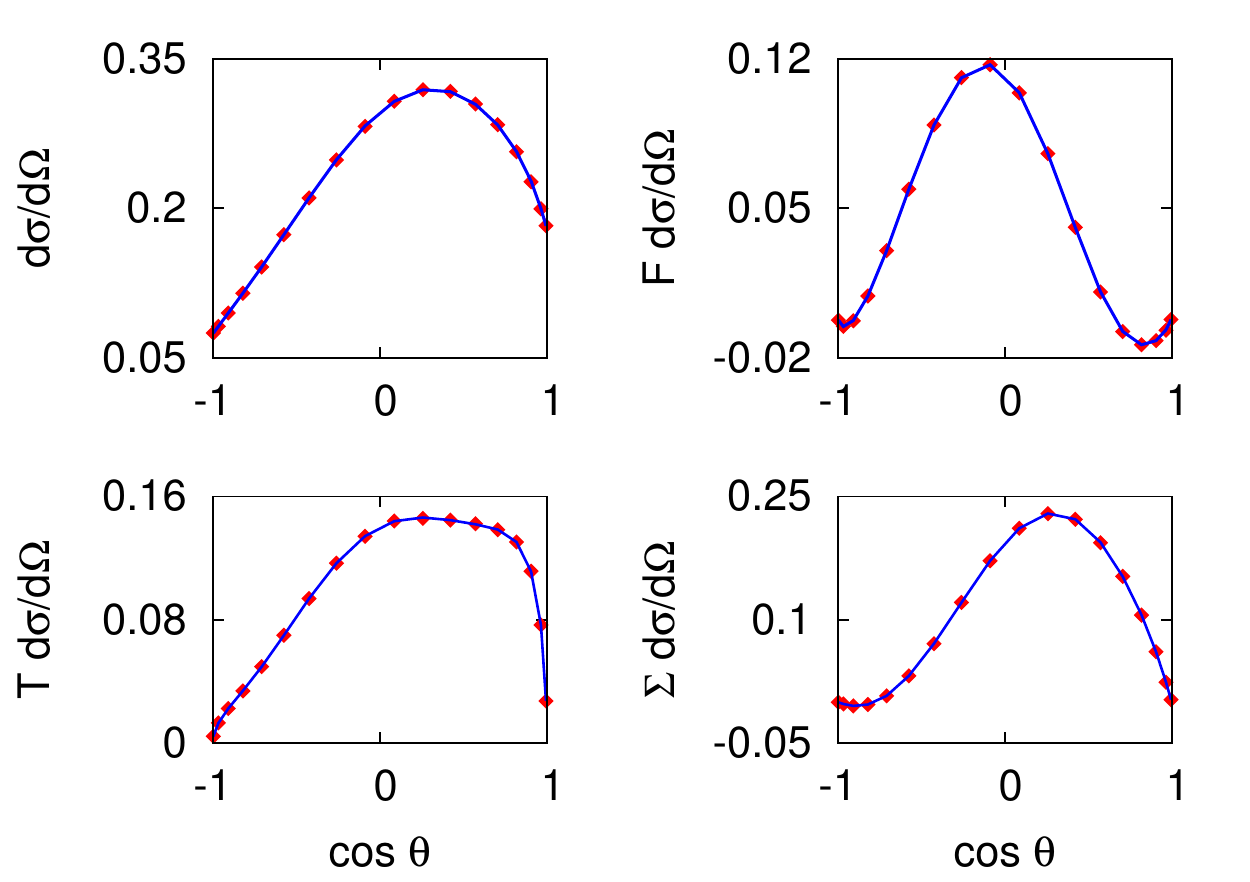}
\includegraphics[width=0.45\textwidth]{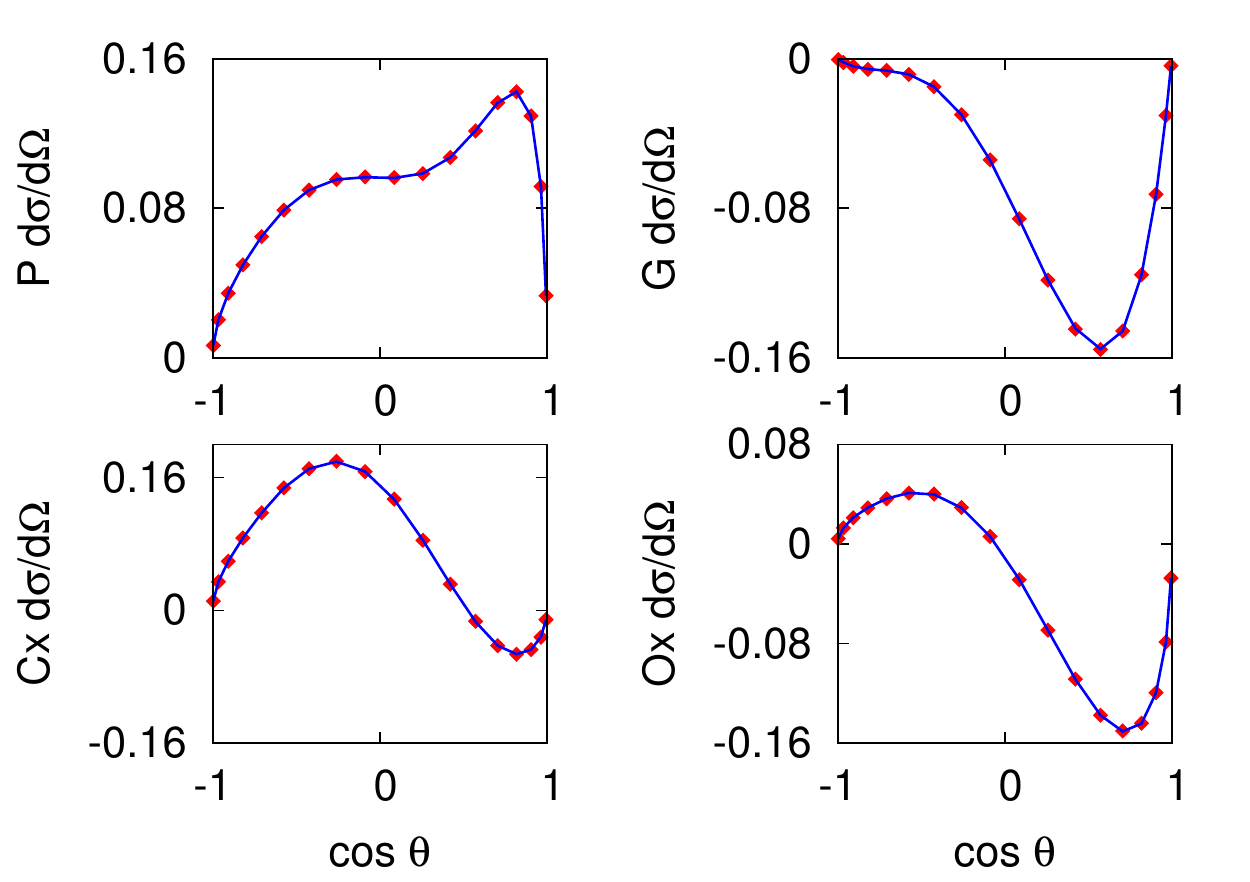}
\end{center}
\vspace*{-0.5cm}
\caption{\label{Observables}(Color online) Complete set of observables for $\eta$ photo-production given in the form of numeric data created at $W= 1769.8$ MeV and at 18 angles using the ETA-MAID15a model (red symbols) and a typical SE fit to the data (full line). }
\end{figure}

We perform unconstrained, $L_{max}=5$ truncated single-energy analyses on a
complete set of observables for $\eta$ photoproduction given in the form of pseudo-data created using the
ETA-MAID15a model
\cite{MAID15a}: $d\sigma /d\Omega$, $ \Sigma \, d\sigma /d\Omega$, $ T \, d\sigma /d\Omega$, $ F \, d\sigma /d\Omega$, $ G \, d\sigma /d\Omega$, $ P \, d\sigma /d\Omega$, $ C_{x'} \, d\sigma /d\Omega$, and $ O_{x'} \,
d\sigma /d\Omega$.   All higher multipoles are put to zero. The fitting procedure finds solutions which are non-unique,
and we obtain many solutions depending on the choice of initial parameters in the fit.  In Fig.~\ref{Observables} we show a complete set of pseudo-data with the error of 1 \% created at 18 angles (red symbols), and the typical SE fit (full line) at one representative energy of $W = 1769.80$ MeV.

\begin{figure}[h!]
\begin{center}
\includegraphics[width=0.75\textwidth]{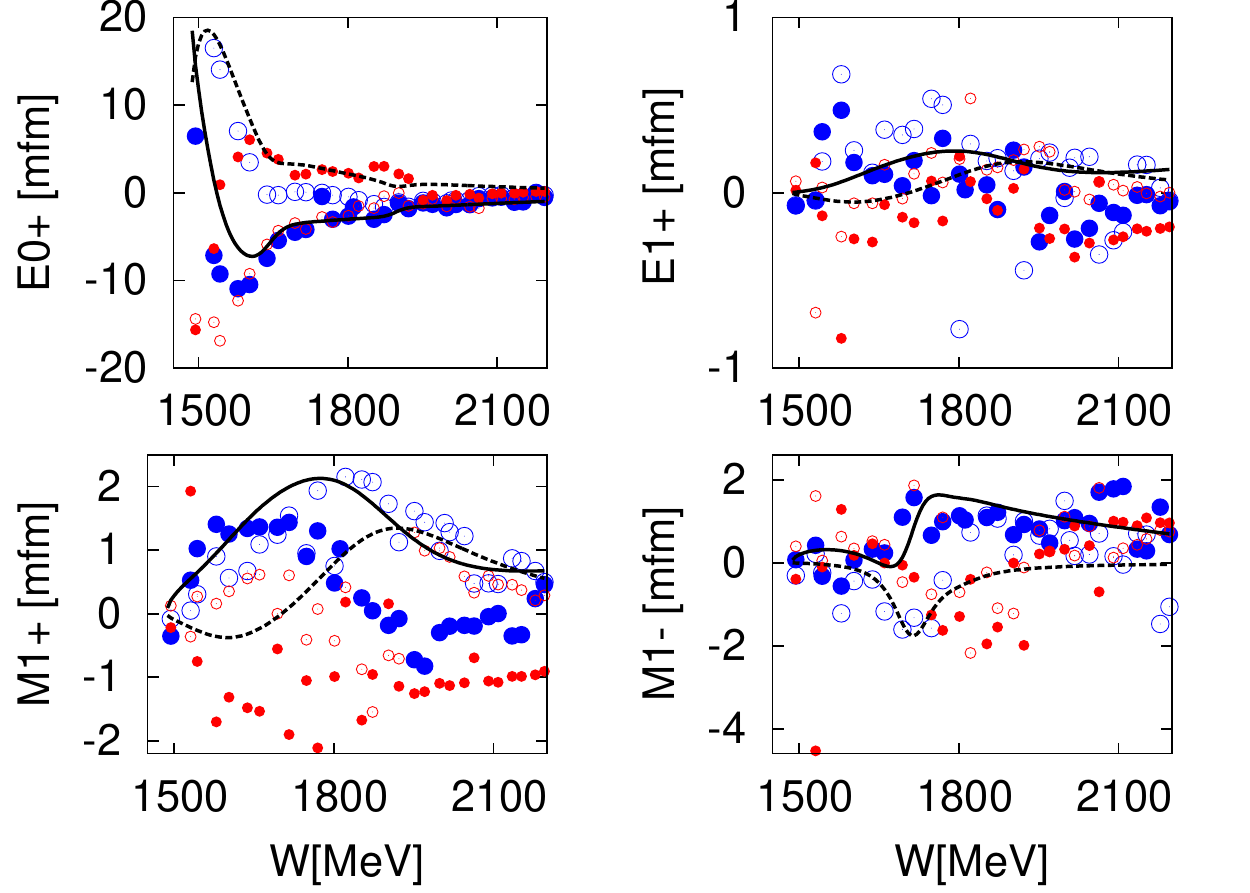}
\end{center}
\vspace*{-0.5cm}
\caption{\label{Fig2}(Color online) Plots of the $E_{0+}$, $M_{1-}$, $E_{1+}$, and $M_{1+}$
multipoles. Full and dashed black lines give the real and imaginary part of the ETA-MAID15a generating model.
Discrete blue and red symbols are obtained  with the unconstrained, $L_{max}=5$  fits of a complete set of
observables  generated as numeric data from the ETA-MAID15a model of ref. \cite{MAID15a},
with the initial fitting values taken from the ETA-MAID16a \cite{MAID16a} and the
Bonn-Gatchina \cite{BoGaweb} models respectively.
Filled symbols represent the real parts and open symbols give the \vspace*{-0.1cm} imaginary parts. }
\end{figure}

In Fig.~\ref{Fig2} we show an example of three very different sets of multipoles which fit the complete pseudo-data set equally well to a high precision:
two discrete and discontinuous ones obtained by setting the initial fitting values to
the ETA-MAID16a \cite{MAID16a} (SE$^{16a}$)  and
Bonn-Gatchina \cite{BoGaweb} (SE$^{BG}$)
model values (blue and red symbols respectively), and the generating ETA-MAID15a model  \cite{MAID15a}
which is displayed as full and  dashed black continuous lines.

\begin{figure}[h!] \hspace*{-0.5cm}
\begin{center}
\includegraphics[width=0.9\textwidth]{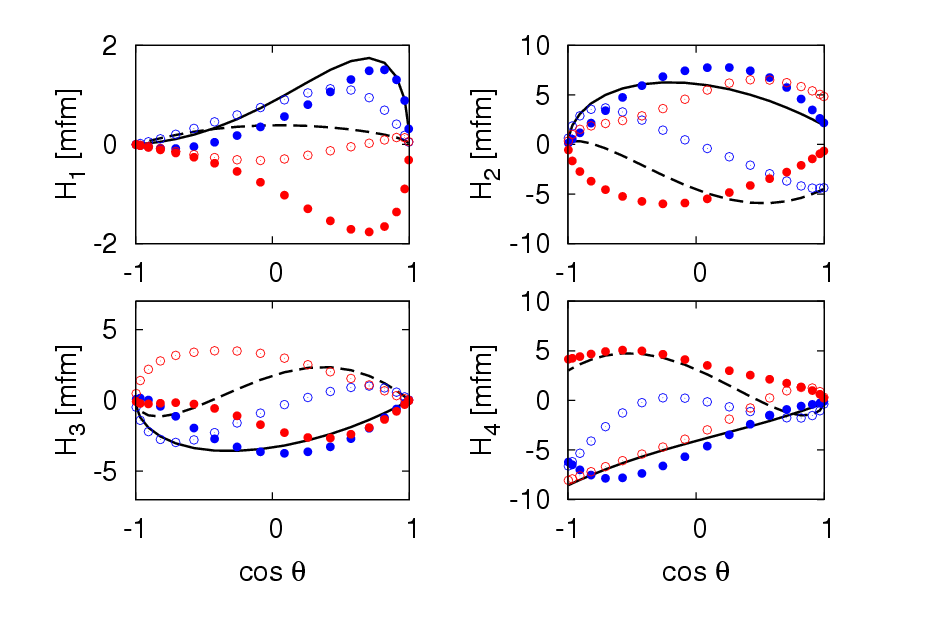}\vspace{-3mm} \\
\caption{\label{Fig3}(Color online)  We show three sets of helicity amplitudes for all three sets of multipoles at one randomly chosen energy  $W= 1660.4$ MeV.
 The figure coding is the same as in Fig.~\ref{Fig2}. }
\end{center}
\end{figure}

\begin{figure}[h!] \hspace*{-0.5cm}
\begin{center}
\includegraphics[width=0.9\textwidth]{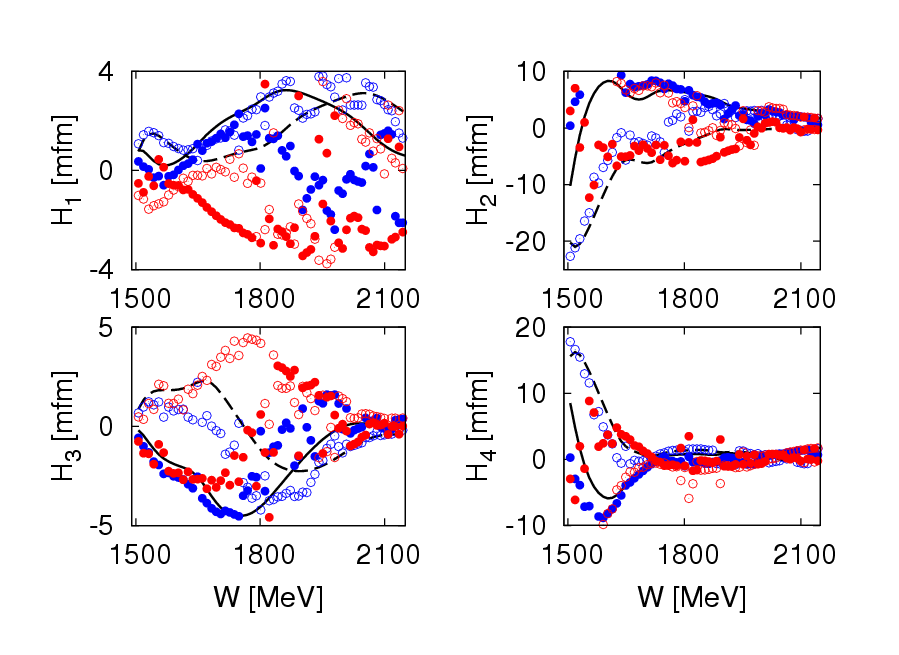}\vspace{-3mm} \\
\caption{\label{Fig3a}(Color online)  Excitation curves of all
helicity amplitudes, for all three sets of multipoles, at one randomly
chosen value of  $\cos \theta= 0.2588$ MeV. The figure coding is the same as in Fig.~\ref{Fig2}.  }
\end{center}
\end{figure}

\begin{figure}[h!]
\begin{center}
\includegraphics[width=0.9\textwidth]{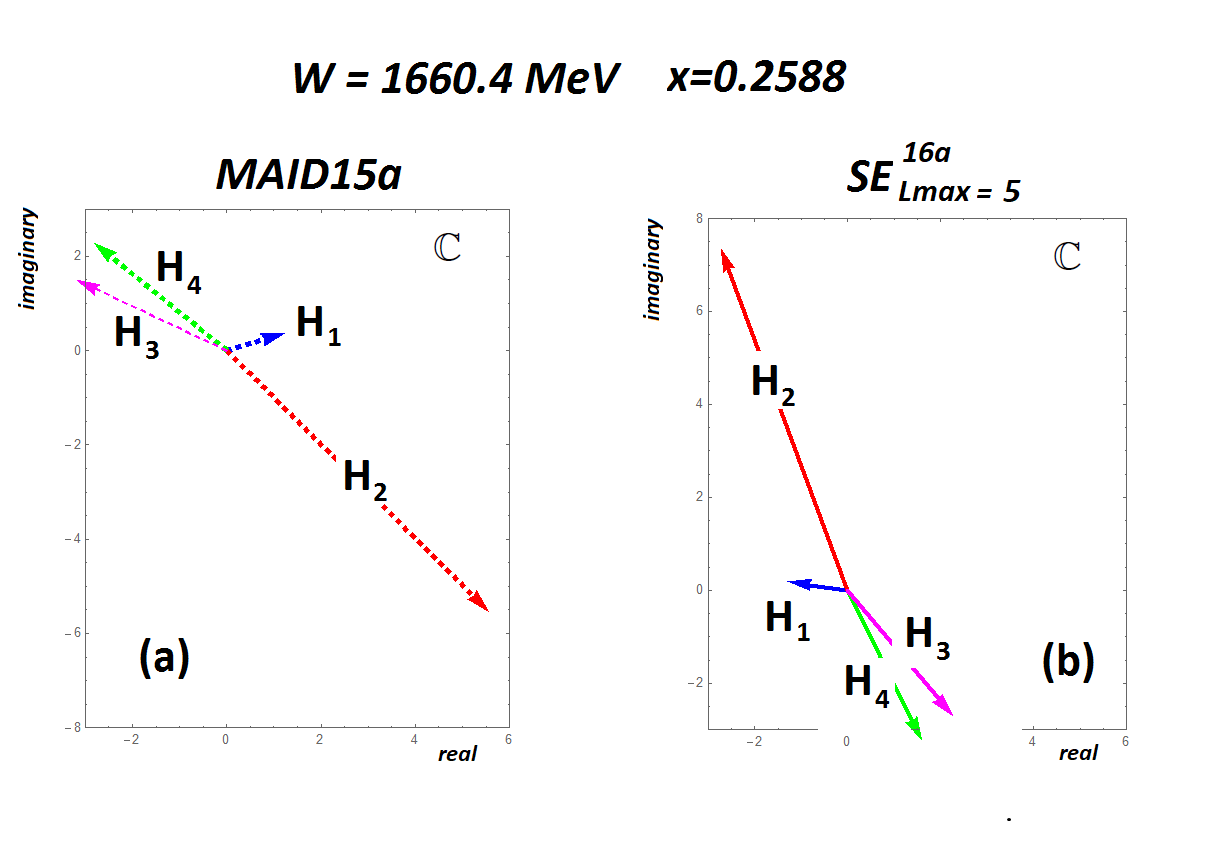}\vspace{-3mm} \\
\caption{\label{Fig4}(Color online) We in Fig.~(a) show all helicity amplitudes for the generating solution MAID15a  and in Fig.~(b) for $SE^{16a}$, $L_{max}=5$ unconstrained solution at one randomly chosen energy of $W= 1660.4$ MeV, and one randomly chosen value
of $\cos \theta=0.2588$. Figures are given in the complex plane $\mathbb{C}$  }
\end{center}
\end{figure}

\begin{figure}[h!]
\includegraphics[width=0.5\textwidth]{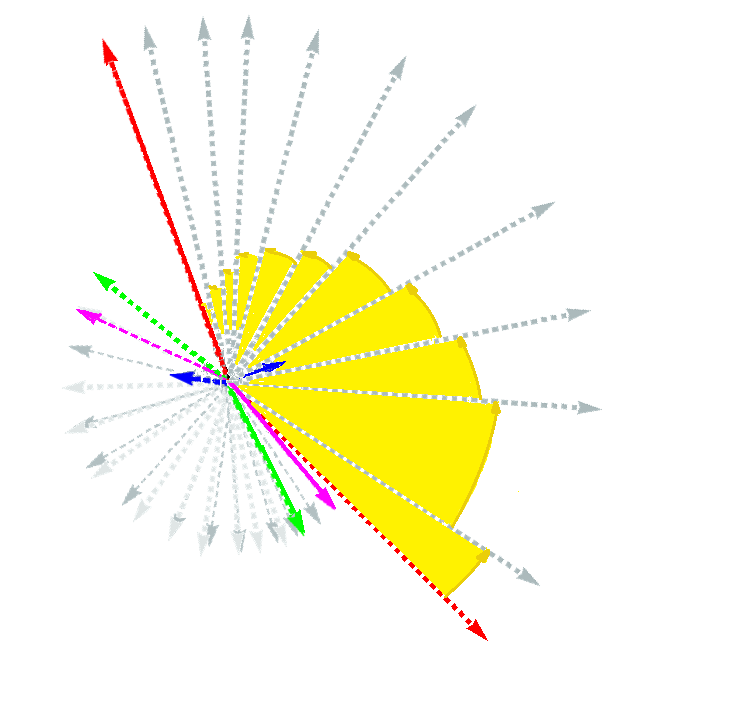}\vspace{-3mm} \\
\caption{\label{Fig4-a}(Color online) We overlap  helicity amplitudes from Figs.~(\ref{Fig4}a) (dashed lines) and (\ref{Fig4}b) (solid lines) at  $W= 1660.4$ MeV and $\cos \theta=0.2588$,  and rotate  them to the same angle in the complex plane. We see that they are explicitly identical after the rotation.  }
\end{figure}

\begin{figure}[h!]
\includegraphics[width=0.8\textwidth]{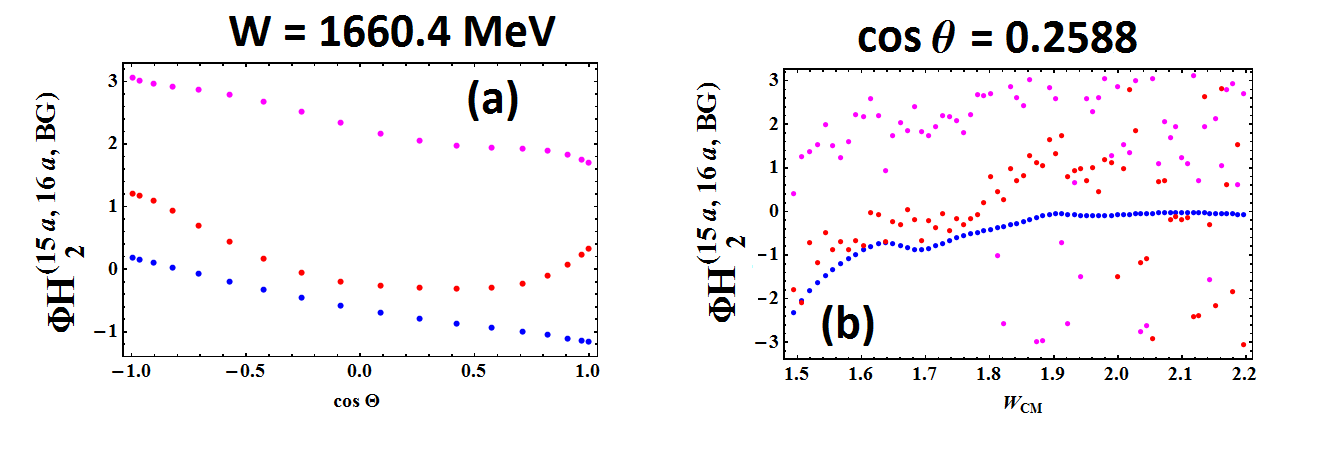}
\includegraphics[width=0.5\textwidth]{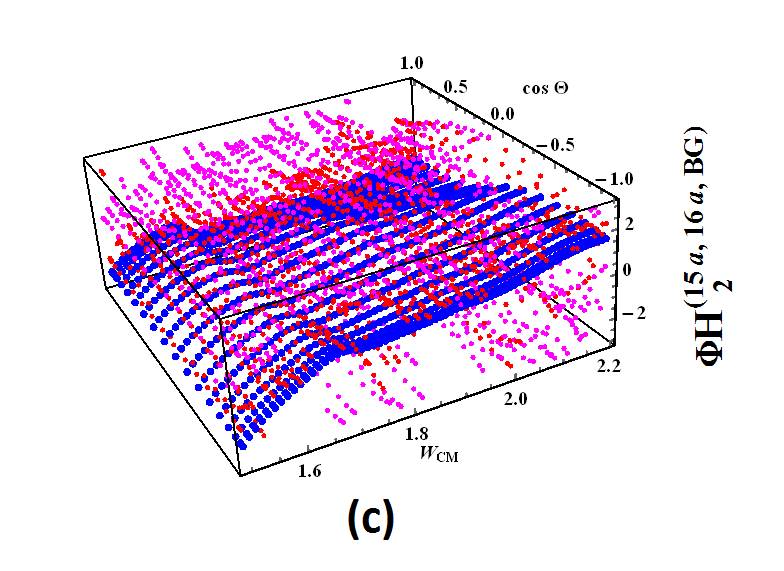}
\caption{\label{Fig-phases}(Color online)  In Fig.~(a) we show the $H_2(W,\theta)$ phase for all three solutions at a randomly chosen energy of  $W= 1660.4$ MeV, and in Fig~(b)  the excitation curve of the same phase at a randomly chosen angle of $\cos \theta=0.2588$. Blue, magenta, and red circles correspond to 15a model,  SE$^{16a}$  and  SE$^{BG}$  solutions respectively. In Fig. (c) we show $H_2(W,\theta)$ phases at all energies and angles in a 3D graph.}
\end{figure}

We know from Eq.(1) that equivalent fits to a complete set of data must be produced by helicity amplitudes with
different phases. Therefore, in Fig.~\ref{Fig3}, we construct the helicity amplitudes
corresponding to all three sets of multipoles from Fig.~\ref{Fig2}
at one randomly chosen energy \mbox{$W= 1660.4$~MeV}.
\\ \\
We see that all three sets of helicity amplitudes are  indeed different,
but the discontinuity of multipole amplitudes, observed in Fig.~\ref{Fig2},
is not reflected in a plot of helicity amplitudes at a fixed single energy.
\newpage
If instead we plot an excitation curve of all four helicity amplitudes
at a randomly chosen angle, which is arbitrarily set to the value  $\cos \theta=0.2588$,
we obtain the result shown in Fig.~\ref{Fig3a}.
\\ \\ We see that the excitation curve of helicity amplitudes in this case
remains continuous only for the generating model ETA-MAID15a.
For both single-energy solutions it is different, and at the same time shows notable discontinuities between
neighbouring energy points.
\newpage To reveal the source of this discontinuity, let us compare helicity amplitudes
at this randomly chosen value of \mbox{$\cos \theta=0.2588$},
at a randomly chosen energy of \mbox{$W= 1660.4$~MeV}.
In Fig.~\ref{Fig4}, the four helicity amplitudes are plotted in the complex plane $\mathbb{C}$ at these arbitrarily chosen energy and angle
for the generating solution and the result of $SE^{16a}$ single-energy fit.
We see that all four helicity amplitudes have identical moduli and relative phases,
but the two sets differ by an overall rotation. In Fig.~\ref{Fig4-a} we demonstrate that for these randomly chosen energy and angle we may rotate one set of helicity amplitudes exactly into another one; so they are identical up to a phase rotation.
\newpage   \noindent
 This effect is valid for each angular point at any arbitrarily chosen energy, so it is valid everywhere. \\ \\ \indent
As  we have shown that there is an obvious mismatch in phases, let us analyse them explicitly. In Fig.~\ref{Fig-phases} we plot the angular and energy dependence of one of the phases, the phase of helicity amplitude $H_2(W,\theta)$. In Fig.~\ref{Fig-phases} (a) we show the $H_2(W,\theta)$ phase for all three solutions at a randomly chosen energy of  $W= 1660.4$ MeV. Blue, magenta, and red circles correspond to 15a model,  SE$^{16a}$  and  SE$^{BG}$  solutions respectively. We see that the phase is continuous for all three solutions.
However, when we in Fig.~\ref{Fig-phases} (b) show the excitation curve of the same phase at a randomly chosen angle of $\cos \theta=0.2588$,  we see that only 15a model phase is continuous; the other two are "jumping" from energy to energy. In Fig.~\ref{Fig-phases} (c) we show $H_2(W,\theta)$ phases at all energies and angles in a 3D graph. We see that phases are continuous at a fixed energy for all three solutions, but for the fixed angle phase stays continuous only for 15a model. For the remaining two it is discontinuous. \\
\newpage
\noindent
This leads to the following understanding of this, apparently very different multipole solutions:
 \\ \\ \indent
\emph{When we perform an unconstrained SE PWA, each minimization is performed independently at individual energies, and the phase is chosen randomly. So, at each energy the fit chooses a different angle dependent phase, and creates different, discontinuous numerical values for each helicity amplitude, producing discontinuous sets of multipoles.  }
\\ \\ \noindent
However, the invariance with respect to phase rotations offers a possible solution.  Let us show the procedure.
\\ \\ \noindent
We introduce the following angle-dependent phase rotation simultaneously for all four helicity amplitudes:
\be \label{Phaserotation}
\tilde{H}_k^{SE}(W, \theta) &=& H_k^{SE}(W, \theta) \, \cdot \, e^{\di \, \Phi_{H_2}^{15a}(W, \theta) - \, \di \, \Phi_{H_2}^{SE}(W, \theta)} \nonumber  \\
k  & = & 1,\ldots,4
\ee
where $\Phi_{H_2}^{SE}(W, \theta)$ is the phase of any single-energy
solution and $\Phi_{H_2}^{15a}(W, \theta)$ is the phase of generating solution ETA-MAID15a.
Applying this rotation we replace the discontinuous $\Phi_{H_2}^{SE} (W, \theta)$ phase from any SE solution with
the continuous $\Phi_{ H_2}^{15a} (W, \theta)$ ETA-MAID15a phase.

\begin{figure}[h!]
\begin{center}
\includegraphics[width=0.84\textwidth]{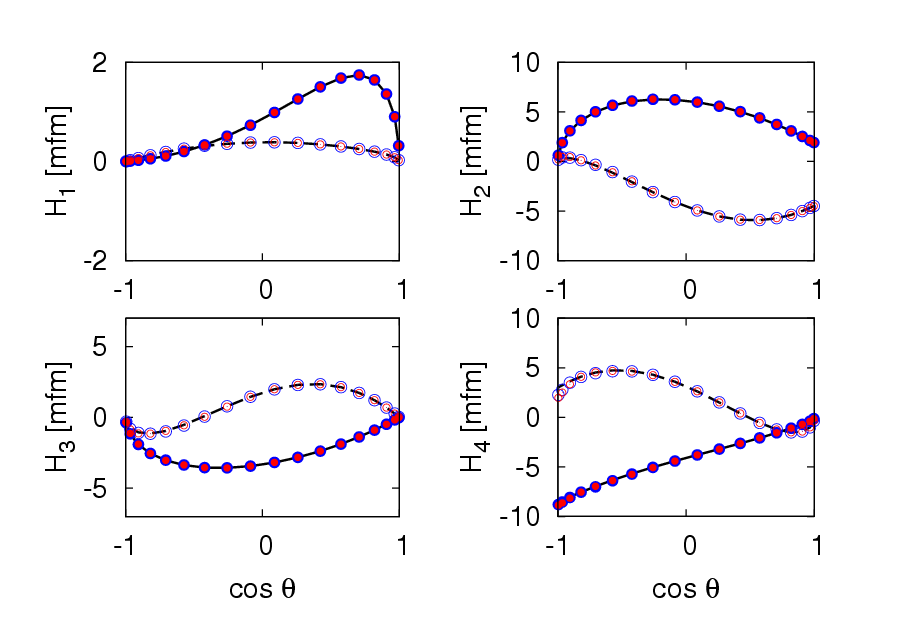}\vspace{-3mm} \\
\caption{\label{Fig5}(Color online) We show all three sets of rotated helicity amplitudes at one randomly chosen energy  $W= 1660.4$ MeV.  The figure coding is the same as in Fig.~\ref{Fig2}. }
\end{center}
\end{figure}

\begin{figure}[h!]
\begin{center}
\includegraphics[width=0.84\textwidth]{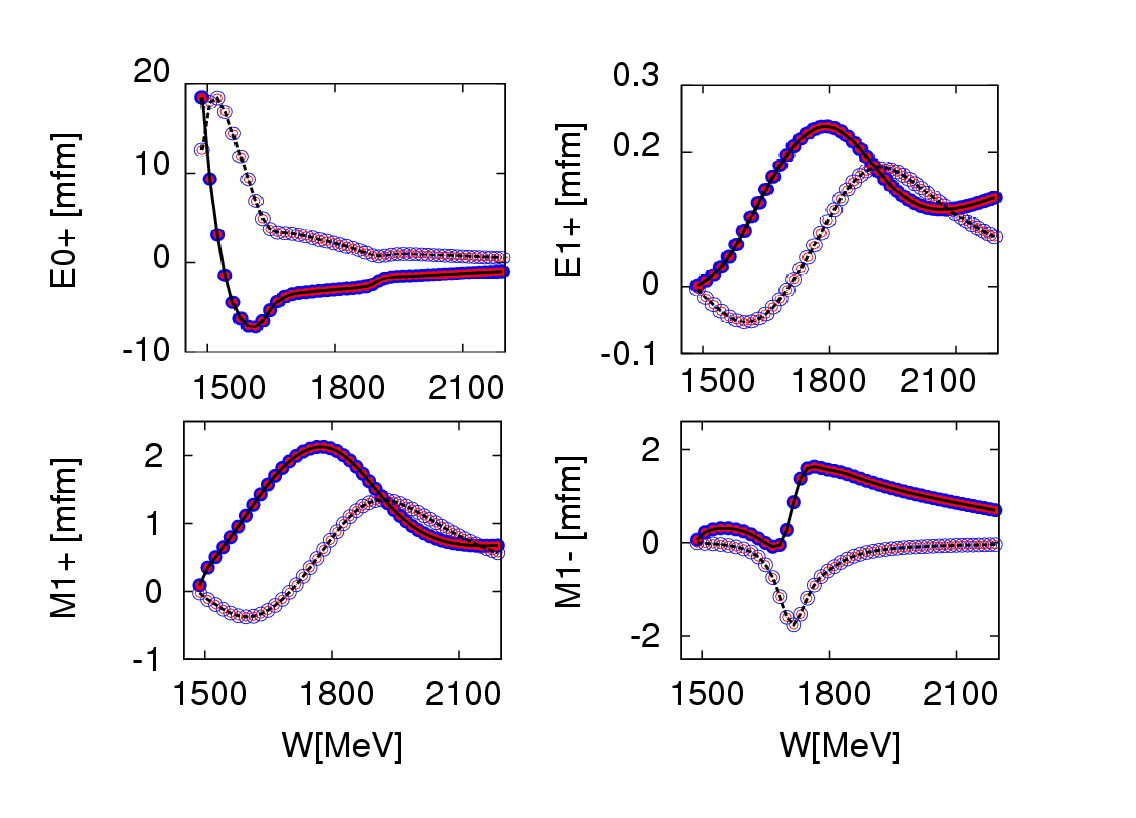}\vspace{-3mm} \\
\caption{\label{Fig6}(Color online)  We show all three sets of rotated multipoles. The figure coding is the same as in Fig.~\ref{Fig2}. }
\end{center}
\end{figure}

The resulting rotated single-energy helicity amplitudes are compared with generating ETA-MAID15a amplitudes
in Fig.~\ref{Fig5}.
\newpage
We see that rotated helicity amplitudes of both single-energy solutions are now identical to the generating ETA-MAID15a
helicity amplitudes.
\\ \\ \indent
Thus, the previously different sets of discrete, discontinuous single-energy
multipoles different from the generating solution ETA-MAID15a and given in Fig.~\ref{Fig2}, are after phase rotation transformed into continuous  multipoles now identical to the generating solution, and given in  Fig.~\ref{Fig6}.
\newpage
 So, we have constructed a way to generate up-to-a-phase unique solutions in an unconstrained PWA of a complete set of observables generated as pseudo-data.
 \\ \\ \indent
 Observe that it is completely inessential which smoothing phase we used in Eq.~(\ref{Phaserotation}) to obtain the unique soluion. In previous tests we used the phase of the generating model $\Phi_{H_2}^{15a}(W, \theta)$ because we knew it in advance, but we could have used any other smooth phase. In that case we would obtain a smooth solution, both SE amplitudes would be the same, but they would differ from the generating ones because of the phase difference with respect to the generating model.
 However, when we rotate it back to the phase of the generating model $\Phi_{H_2}^{15a}(W, \theta)$, the input quantities are again reproduced. Here we give the example if the smoothing phase is chosen to be the phase of the Toy model  $\mathcal{R}(x)$.
 \\ \\ \indent
 Instead of rotating the SE reaction amplitudes with a known phase of the generating model, we rotate it with any arbitrary, but continuous phase. For simplicity, we use the same phase $\mathcal{R}(x)$ which we used for the toy model, so our phase rotation previously given in  Eq.~(\ref{Phaserotation}) now reads:
 \be \label{Phaserotation-new}
\tilde{H}_k^{SE}(W, \theta) &=& H_k^{SE}(W, \theta) \, \cdot \, e^{\di \, (2. + 0.5 \, x)  - \, \di \, \Phi_{H_2}^{SE}(W, \theta)} \nonumber  \\
k  & = & 1,\ldots,4
\ee
\newpage
We obtain smooth helicity amplitudes at a fixed energy  $W= 1660.4$ MeV:
\begin{figure}[h!]
\begin{center}
\includegraphics[width=0.8\textwidth,height=0.39\textheight]{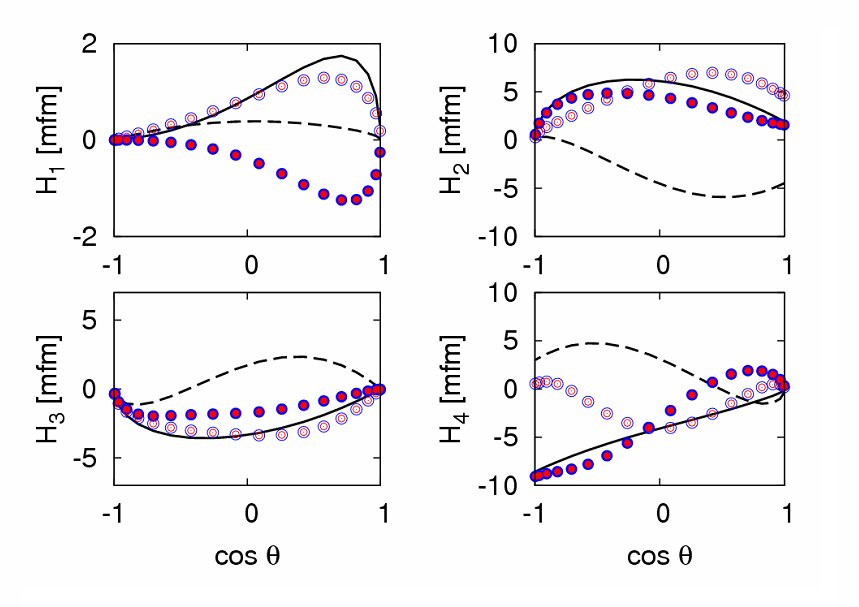}\vspace{-3mm} \\
\caption{\label{Arbitrary-phase-W-fixed}(Color online)  We show all three sets of helicity amplitudes rotated with Toy model rotation $\mathcal{R}(x)$ at one randomly chosen energy  $W= 1660.4$ MeV.  The figure coding is the same as in Fig.~\ref{Fig2}. }
\end{center}
\end{figure}
\\
However, contrary to unconstrained case, we also obtain a smooth behaviour for the excitation curve with the fixed angle corresponding to  $\cos \theta=0.2588$:
\begin{figure}[h!]
\begin{center}
\includegraphics[width=0.8\textwidth,,height=0.39\textheight]{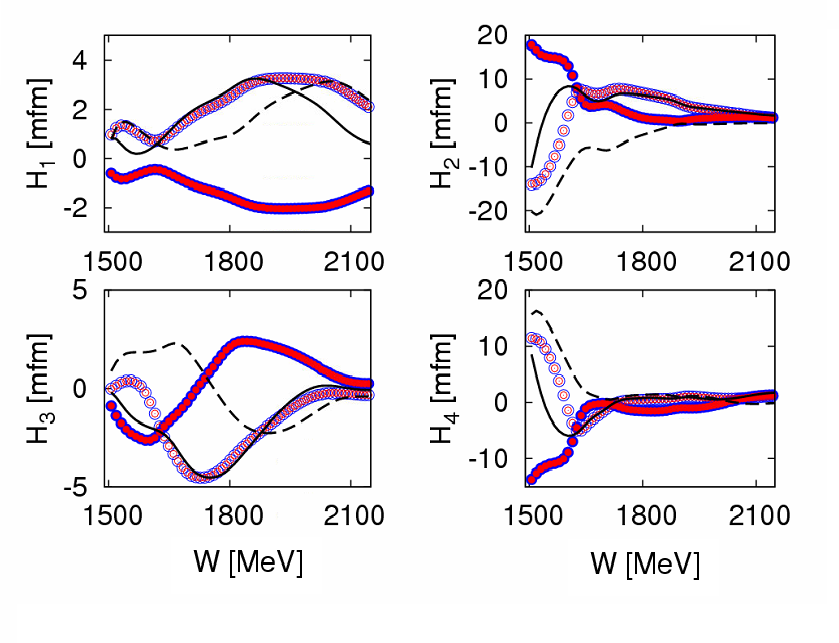}\vspace{-3mm} \\
\caption{\label{Arbitrary-phase-theta-fixed}(Color online) We show all three sets of helicity amplitudes rotated with Toy model rotation $\mathcal{R}(x)$ at one randomly chosen angle which corresponds to $\cos \theta=0.2588$.  The figure coding is the same as in Fig.~\ref{Fig2}. }
\end{center}
\end{figure}
\newpage \noindent

Observe that both rotated curves are now smooth and identical, but different from the generating model values. So, we have achieved the uniqueness. By replacing the discontinuous phase with an arbitrary, but continuous one we achieved that our solutions do not depend on the choice of initial values. We still have to achieve that they reproduce the generating value.
\\ \\ \indent
However, if we apply the rotation which transforms $\mathcal{R}(x) \rightarrow  \Phi_{H_2}^{15a}(W, \theta)$:
\be \label{Phaserotation-new}
\tilde{H}_k^{SE}(W, \theta) &=& H_k^{SE}(W, \theta) \, \cdot \, e^{\di \, \Phi_{H_2}^{15a}(W, \theta) - \, \di \,  (2. + 0.5 \, x)} \nonumber  \\
k  & = & 1,\ldots,4
\ee
we reproduce values of the generating model given in Figs.~\ref{Fig5}~and~\ref{Fig6}.
 \\ \\ \indent
 So, smoothing discontinuous solutions shown in Fig.~\ref{Fig-phases}, and not the choice of smoothing phase is essential. Once we get a smooth solution, it is unique in a sense that we can rotate it to any other phase of our choice without any further minimizations. We have obtained an up-to-a-phase unique solution. Determining the proper phase is, however, essential because without it a correct association of quantum  numbers is impossible.
 \section{Laurent + Pietarinen expansion}
The driving concept behind the Laurent-Pietarinen (L+P) expansion was the aim to replace
an elaborate theoretical model by a local power-series representation of partial wave
amplitudes~\cite{Svarc2013}. The complexity of a partial-wave analysis model
is thus replaced by much simpler model-independent expansion which just exploits analyticity
and unitarity. The L+P approach separates pole and regular part in the form of a Laurent expansion, and instead of modeling the regular part in some physical model it uses the conformal mapping to expand
it into a rapidly converging power series with well defined analytic properties. So, the method replaces the regular part calculated in a model by the simplest analytic function which has correct analytic properties of the analyzed partial wave (multipole), and fits the data. In such an approach the model dependence is minimized, and is reduced to the choice of the number and location of branch-points used in the model.

The L+P expansion is based on the Pietarinen expansion used in some former papers in the analysis of pion-nucleon scattering data
\cite{Ciulli,CiulliFisher,Pietarinen,Pietarinen1}, but for the L+P model the Pietarinen expansion is applied in a different manner. It exploits the Mittag-Leffler expansion\footnote{Mittag-Leffler expansion \cite{Mittag-Leffler}. This expansion is the generalization of a Laurent expansion to a
more-than-one pole situation. For simplicity, we will simply refer to this as a Laurent expansion.} of partial wave amplitudes near the real energy axis, representing the regular, but unknown, background term by a conformal-mapping-gener\-ated, rapidly converging power series called a Pietarinen expansion\footnote{A conformal mapping expansion of this particular type was introduced by Ciulli and Fisher \cite{Ciulli,CiulliFisher}, was described in detail and used in pion-nucleon scattering by Esco Pietarinen \cite{Pietarinen,Pietarinen1}. The procedure was denoted as a Pietarinen expansion by G. H\"{o}hler in
\cite{Hohler:1984ux}.}.  The method was used successfully in several
few-body reactions \cite{Svarc2014,Svarc2014a,L+P2015}, and recently generalized to the multi-channel
case \cite{Svarc2016}. The formulae used in the L+P approach are collected in Table~\ref{LplusP}.
\begin{table*}[h!]
\caption{\label{LplusP}Formulae defining the Laurent+Pietarinen (L+P) expansion.\vspace{-5mm}}
{\footnotesize
\begin{align*}
\label{eq:Laurent-Pietarinen}
T^a(W) &=& \sum _{j=1}^{{N}_{pole}} \frac{x^{a}_{j} + \imath \, \, y^{a}_{j}  }{W_j-W} +
      \sum _{k=0}^{K^a}  c^a_k \, X^a (W)^k  +  \sum _{l=0}^{L^a} d_l^a \, Y^a (W)^l +  \sum
_{m=0}^{M^a} e_m^a \, Z^a (W)^m \nonumber\hspace{50mm} \\
X^a (W )&=&  \frac{\alpha^a-\sqrt{x_P^a-W}}{\alpha^a+\sqrt{x_P^a - W }}; \, \, \, \, \,   Y^a(W ) =  \frac{\beta^a-\sqrt{x_Q^a-W }}{\beta^a+\sqrt{x_Q^a-W }};  \, \, \, \, \,
Z^a(W ) =  \frac{\gamma^a-\sqrt{x_R^a-W}}{\gamma^a+\sqrt{x_R^a-W }}
 \nonumber \hspace{42mm}\\
  D_{dp}^a \ \ \ \ \ & = & \frac{1}{2 \, N_{W}^a - N_{par}^a} \, \, \sum_{i=1}^{N_{W}^a} \left\{ \left[ \frac{{\rm Re} \,T^{a}(W^{(i)})-{\rm Re} \, T^{a,exp}(W^{(i)})}{ Err_{i,a}^{\rm Re}}  \right]^2 + \right.
            \left. \left[ \frac{{\rm Im} \, T^{a}(W^{(i)})-{\rm Im} \, T^{a,exp}(W^{(i)})}{ Err_{i,a}^{\rm Im}} \right]^2 \right\}  + {\cal P}^a  \nonumber\hspace{3mm} \\
{\cal P}^{a}\ \ \ \ \  &=& \lambda_c^a \, \sum _{k=1}^{K^a} (c^a_k)^2 \, k^3 +
\lambda_d^a \, \sum _{l=1}^{L^a} (d^a_l)^2 \, l^3 +  \lambda_e^a \,
\sum _{m=1}^{M^a} (e^a_m)^2 \, m^3 \qquad\qquad
 D_{dp}  =  \sum _{a}^{all}D_{dp}^a \nonumber \hspace{40mm}\\
 &&a\, \, .....   \, \,    {\rm  channel \, \,  index} \qquad\qquad
  N_{pole}\; .....\;      {\rm number\; of\; poles} \qquad\qquad
  W_j,W \in \mathbb{C}  \nonumber \hspace{42mm}\\
&& x_i^a, \, y_i^a, \, c_k^a, \, d_l^a, \, e_m^a, \, \alpha^a, \, \beta^a, \, \gamma^a ... \in  \mathbb{R}  \, \,
 \nonumber  \hspace{96mm}\\
 &&  K^a, \, L^a, \, M^a \, ... \, \in  \mathbb{N} \, \, \, {\rm number \, \, of \, \, Pietarinen \, \, coefficients \, \, in \, \, channel \, \, \mathit{a} }.
 \nonumber \hspace{53mm}\\
 && D_{dp}^a \; .....  \; {\rm discrepancy \; function \; in\; channel \; }a \qquad\qquad \hspace{12mm}
N_{W}^a \; .....  \; {\rm number\; of\; energies\; in\; channel \; }a \nonumber \hspace{8mm}\\
 && N_{par}^a \; .....  \; {\rm number\; of\; fitting \; parameters \; in\; channel \; }a
\qquad\qquad {\cal P}^a \, \,  .....   \, \, {\rm Pietarinen  \, \, penalty \, \, function}
\nonumber \hspace{14mm}\\
 && \lambda_c^a, \, \lambda_d^a, \,\lambda_e^a \, \, \,  .....   \, \, {\rm Pietarinen  \, \,  weighting \, \, factors} \nonumber \qquad\qquad\hspace{10mm} x_P^a, \, x_Q^a, \, x_R^a  \in \mathbb{R}
 \, \, \, \,( {\rm  or} \, \, \in \mathbb{C}). \nonumber  \hspace{26mm}
\\
&& Err_{i,a}^{\rm Re, \, Im} ..... {\rm \, \, minimization \, \, error\, \, of \, \, real \,\, and \, \, imaginary \, \, part \, \, respectively.} \nonumber \hspace{45mm}
\end{align*}
}%
\end{table*}
In the fits, the regular background part is represented by three Pietarinen expansion series, all
free parameters are fitted. The first Pietarinen expansion with branch-point $x_P$ is restricted
to an unphysical energy range and represents all left-hand cut contributions. The next two Pietarinen
expansions describe the background in the physical range with branch-points $x_Q$ and $x_R$ respecting the analytic properties of the analyzed partial wave. The second branch-point is mostly fixed to the
elastic channel branch-point, the third one is either fixed to the dominant channel threshold,
or left free.  Thus, only rather general physical assumptions about the analytic properties are made
like the number of poles and the number and the position of branch-points, and the simplest analytic
function with a set of poles and branch-points is constructed. The method is applicable to both,
theoretical and experimental
input, and represents the first reliable procedure to extract pole positions from experimental data, with minimal model bias.

The generalization of the L+P method to a multichannel L+P method is performed in the following way:
i)~separate Laurent expansions are made for each channel; ii)~pole positions are fixed for all
channels, iii)~residua and Pietarinen coefficients are varied freely; iv)~branch-points are chosen
as for the single-channel model; v)~the single-channel discrepancy function $D_{dp}^a$ (see Eq. (5) in ref. \cite{L+P2015}) which quantifies the deviation of the fitted function from the input is generalized
to a multi-channel quantity $D_{dp}$ by summing up all single-channel contributions, and vi)~the minimization is performed for all channels in order to obtain the final solution.

The formulae used in the L+P approach are collected in Table~\ref{LplusP}.

\section{Conclusions}
Let us summarize by affirming that taking into consideration the angle dependent continuum ambiguity phase is not only an academic issue; it is essential. The problem it introduces is that it mixes partial waves and hence modifies resonance quantum number association. On the other hand, taking the benefit of it enables one to obtain an up-to-a phase unique solution in the unconstrained PWA of a complete set of observables.

However, in practical, day-to-day calculations the continuum ambiguity must, in some way, be taken into account to actually
obtain a continuous partial-wave solution. Without fixing the overall angle and energy dependent phase,
a whole class of equivalent reaction amplitude sets, connected via rotations exist, and that produces discontinuities in unconstrained SE PWA.  We eliminate this problem by fixing the unknown phase to any continuous one. In this way a solution with a continuous, but predefined phase is obtained. However, it  by no means represents the loss of generality. Namely, once we have obtained a unique solution with a known phase, we can without any further minimizations simply rotate this solution to any other phase (we reconstruct the helicity amplitudes, use Eq.~(\ref{Phaserotation}) to change the phase to any desired value, and project onto partial waves), and obtain a partial wave decomposition belonging to this new phase.   It is also important to keep in mind that these rotations do not create new poles but just reshuffles the old ones. However, this may put poles into partial waves having different quantum numbers. Unfortunately this means that resonance quantum numbers may be altered, so at the end we have to determine the exact continuum ambiguity phase to correctly assign quantum numbers to the established resonances.

Let us also state that the continuum ambiguity phase is not just a mathematical fiction without possibility to be determined. First, it definitely exists, and seriously influences our reasoning. Second, it can also be exactly determined; definitely in coupled-channel calculations.  When only one channel is analyzed, unitarity constraints relating real and imaginary parts of the partial wave T-matrices are lost after the first inelastic threshold opens so continuum ambiguities arises, but it is automatically restored when all channels are included in coupled-channel calculations (uncontrolled loss of flux in one channel is controlled by including all of them). So, coupled-channel formalisms are the natural way to eliminate continuum ambiguities.

However, even when we managed to obtain "up-to-a-phase" unique solution,  there was no reliable way to extract pole parameters from so obtained SE partial waves, but a new and simple single-channel method (Laurent + Pietarinen expansion) applicable for continuous and discrete data has been recently developed. It is based on applying the Laurent decomposition of partial wave amplitude, and expanding the non-resonant background into a power series of a conformal-mapping, quickly converging power series obtaining the simplest analytic function with well-defined partial wave analytic properties which fits the input. The method is particularly useful to analyse partial wave data obtained directly from experiment because it works with minimal theoretical bias since it avoids constructing and solving elaborate theoretical models, and fitting the final parameters to the input, what is the standard procedure now.  The generalization of this method to multi- channel case is also developed and presented.

 Unifying both methods in succession, one constructs a model independent procedure to extract pole parameters directly from experimental data without referring to any theoretical model.

\section*{Acknowledgements}
This work was supported by the Deutsche Forschungsgemeinschaft (SFB TR16 and 1044). The work of
RW was supported by the U.S. Department of Energy grant DE-SC0016582.


\clearpage

\begin{thebibliography}{AA}
\bibitem{MartinSpearman} A.D. Martin and T.D. Spearman: Elementary Particle Theory, North-Holland Publishing Company, Amsterdam 1970.
\bibitem{Atk73} D. Atkinson, P.W. Johnson and R.L. Warnock, Commun. mat. Phys. {\bf 33} (1973) 221.
\bibitem{Bow75} J.E. Bowcock and H. Burkhard, Rep. Prog. Phys. {\bf 38} (1975) 1099.
\bibitem{Atk85} D. Atkinson and I.S. Stefanescu, Commun. Math. Phys. {\bf 101}, 291 (1985).
\bibitem{bnga}                                                                                                                                                                                                                                                                                                                                                                              A.V. Anisovich, R. Beck, E. Klempt, V.A. Nikonov, A.V. Sarantsev, U. Thoma,
Eur. Phys. J. \textbf{A48}, 15 (2012).
\bibitem{Tiator:2011pw}
  L.~Tiator, D.~Drechsel, S.~S.~Kamalov and M.~Vanderhaeghen,
  Eur.\ Phys.\ J.\ ST {\bf 198}, 141 (2011).
\bibitem{Sandorfi2011} A. M. Sandorfi, S. Hoblit, H. Kamano,  and T.-S. H. Lee, J. Phys. G: Nucl. Part. Phys. {\bf 38} (2011) 053001.
\bibitem{Oma81} A.S. Omalaenko, Sov. Jour. Nucl. Phys. \textbf{34}(3) (1981) 406-411.
\bibitem{Dean} N.W. Dean and P. Lee, Phys. Rev. D \textbf{5}, 2741 (1972).
\bibitem{Keaton} G. Keaton and R. Workman, Phys. Rev. C \textbf{54}, 1437 (1996).
\bibitem{Svarc2013} A. Svarc, M. Hadzimehmedovic, H. Osmanovic, J. Stahov, L. Tiator, and R. L. Workman, Phys, 	Rev.  \textbf{ 	C88}, 035206 (2013).
\bibitem{Svarc2014} A. Svarc,  M. Hadzimehmedovic, R. Omerovic, H. Osmanovic, and J. Stahov,  Phys, Rev. \textbf{C89},   	0452205  (2014).
\bibitem{Svarc2014a}   A. Svarc, M. Hadzimehmedovic, H. Osmanovic, J. Stahov, L. Tiator, and R. L. Workman, Phys, 	Rev.  \textbf{C89}, 65208 (2014).
\bibitem{Dougall1952} J. Dougall, Glasgow Mathematical Journal, {\bf 1} (1952) 121-125.
\bibitem{Wunderlich2017} Y. Wunderlich, A. \v{S}varc, R. L. Workman, L. Tiator, and R. Beck,  arXiv:1708.06840[nucl-th].
\bibitem{Pseudo-data} see R. L. Workman, M. W. Paris, W. J. Briscoe, L. Tiator, S. Schumann, M. Ostrick, S. S. Kamalov, EPJA (2011) \textbf{47}:143, \emph{and references therein}.
\bibitem{MAID15a} V. L. Kashevarov, L. Tiator, M. Ostrick, Bled Workshops Phys., \textbf{16}, 9 (2015).
\bibitem{MAID16a} V. L. Kashevarov, l. Tiator, M. Ostrick, JPS Conf. Proc. \textbf{13}, 020029 (2017).
\bibitem{BoGaweb} http:\-/\-/pwa.hiskp.uni-bonn.de\-/
\bibitem{TPWAMainz} R.L. Workman, L. Tiator, Y. Wunderlich, M. Doering, H. Haberzettl, Phys. Rev. C 95, 015206 (2017).
\bibitem{Regge}  J. Nys, V. Mathieu, C. Fernández-Ramírez, A. N. Hiller Blin, A. Jackura, M. Mikhasenko, A. Pilloni, A. P. Szczepaniak, G. Fox, J. Ryckebusch,
Phys. Rev. \textbf{D 95}, 034014 (2017).
\bibitem{SE_test} R.L. Workman, M.W. Paris, W.J. Briscoe, L. Tiator, S. Schumann, M. Ostrick, and S.S. Kamalov, Eur. Phys. J. A \textbf{47}, 143 (2011).
\bibitem{Ciulli}S. Ciulli and J. Fischer in Nucl. Phys. \textbf{24}, 465 (1961).
\bibitem{CiulliFisher}I. Ciulli, S. Ciulli, and J. Fisher, Nuovo Cimento \textbf{23}, 1129 (1962).
\bibitem{Pietarinen} E. Pietarinen, Nuovo Cimento Soc. Ital. Fis. \textbf{12A}, 522 (1972).
\bibitem{Pietarinen1} E. Pietarinen, Nucl. Phys. \textbf{B107}, 21 (1976).
\bibitem{Mittag-Leffler}Michiel Hazewinkel: \emph{Encyclopaedia of Mathematics}, Vol.6,  Springer, 31. 8. 1990, pg.251.
\bibitem{Hohler:1984ux} G.~H\"ohler and H.~Schopper,
``Numerical Data And Functional Relationships In Science And
Technology. Group I: Nuclear And Particle Physics. Vol. 9: Elastic
And Charge Exchange Scattering Of Elementary Particles. B: Pion
Nucleon Scattering. Pt. 2: Methods And Results and Phenomenology,'' {\it  Berlin, Germany:
Springer ( 1983) 601 P. ( Landolt-Boernstein. New Series, I/9B2).}
\bibitem{Svarc2016} A. \v{Svarc}, M. Had\v{z}imehmedovi\'{c}, H. Osmanovi\'{c}, J. Stahov, L. Tiator, R. L. Workman,
 Phys. Lett. \textbf{B755} (2016) 452–455.
\bibitem{L+P2015} A. \v{Svarc}, M. Had\v{z}imehmedovi\'{c}, H. Osmanovi\'{c}, J. Stahov,  and R. L. Workman, Phys. Rev. \textbf{C91}, {015207} (2015).


\end{thebibliography}
\end{document}